# Minimum Quench Energy of Nb₃Sn Wires with High Specific Heat Tape

E. Barzi, *Senior Member, IEEE*, I. Novitsky, D. Turrioni, A. V. Zlobin, X. Peng, M. Tomsic

*Abstract*— A major problem of state-of-the-art Nb₃Sn accelerator magnets is their long training due to thermo-mechanical perturbations. Increasing the specific heat, $C_p$, of the Rutherford cable would reduce and/or eliminate training by limiting the coils temperature rise. This paper studies feasibility of increasing the $C_p$ of Rutherford-type cables by using thin composite Cu/Gd₂O₃ and Cu/Gd₂O₂S tapes produced by Hyper Tech Research, Inc. The tape can be either wrapped around the cable, placed on the cable wide faces under the insulation, and/or inserted as a core. Wire samples outfitted with these high-$C_p$ ribbons, or tapes, were prepared and tested at FNAL for their Minimum Quench Energy (*MQE*). At 90%$I_c$ and 15 T, the average gain of *MQE* of the Nb₃Sn wire soldered to the Cu/Gd₂O₂S 55 μm thick ribbon was 2.5, and further increased at larger transport currents.

## I. Introduction

A remaining problem in Nb₃Sn high field accelerator magnets R&D for high energy physics (HEP) and other applications is their long training [1], [2]. Superconducting (SC) magnets quench when their temperature increases above the current sharing temperature of the composite superconductor over a large enough volume. The temperature increase $\Delta T$ is proportional to $Q/C_p$, where $Q$ is the dissipated heat, and $C_p$ is the volumetric heat capacity. Sources of magnet training include magnetic flux jumps, conductor motion, epoxy cracking, and material interfaces, such as between impregnating material and neighboring structural materials for instance. All these sources result in a "disturbance spectrum".

Perturbations may originate either outside or inside the superconductor, and they can be characterized by three parameters: spatial distribution, duration and energy or power amplitudes [3]. Normal zone propagation is slowed down by heat diffusion from the normal region and heat transfer to the coolant. Depending on the thermal disturbance energy, the normal zone either disappears or grows infinitely. A stationary temperature profile exists for which the Joule power is in equilibrium with the power extracted by conduction and by the external environment. This stationary solution is designated as Minimum Propagating Zone (*MPZ*). Up until the late 70s, it was generally accepted that an instantaneous perturbation, described by a localized external input of heat $Q_0$, led to propagation of the normal zone if $Q_0$ exceeded the Minimum Quench Energy (*MQE*), calculated as the stationary enthalpy of the *MPZ* [4], [5]. However, this theory leads to an estimate of the *MQE* only for distributed disturbances, i.e. on the order of the characteristic thermal length, or $(kA/hp)^{1/2}$, where $k$ is the conductor thermal conductivity, $A$ its cross-section area, $h$ the convective heat exchange coefficient and $p$ the cooled perimeter [6], [7]. For the Rutherford-type cables in this study, the characteristic thermal length ranges from 3 mm to 2.5 cm with heat transfer coefficients $h$ respectively from 1 W/(cm²K), which may apply to Kapton insulated NbTi cable in boiling Helium, to 0.01 W/(cm²K), which may apply to impregnated Nb₃Sn cable.

The *MQE* of a local heat pulse may be higher or lower than the distributed *MPZ* energy. It was shown in [8] that this depends on the fraction of transport current $I$ with respect to the critical current $I_c$. In the $I$ region close to $I_c$, the *MQE* is smaller than the *MPZ* energy, and in the region of smaller current fractions it is larger. The boundary between the two regions depends on the cooling conditions. The worse is the cooling, the closer the boundary current is to zero. The nucleate boiling conditions of the cable samples used in the present study place the boundary value at ~50% $I_c$.

The idea to increase the *MQE* by inserting high specific heat (high-$C_p$) elements in superconducting wires dates to 1960 [9]. More recently, Hyper Tech and Bruker-OST have attempted to introduce high-$C_p$ elements in their wire design [10]. An alternate approach is to introduce high-$C_p$ components, such as an high-$C_p$ ribbon, in the Rutherford cable itself [2]. Hyper Tech produced high-$C_p$ ribbons by rolling down to various thicknesses the Cu/Gd₂O₃ and Cu/Gd₂O₂S high-$C_p$ tubes used for their wires. In [2], the positive effect on stability was measured for NbTi wires and Rutherford cables outfitted with Cu/Gd₂O₃ tapes. In this study, samples of both Cu/Gd₂O₃ and Cu/Gd₂O₂S tapes were used to measure their effect on stability of Nb₃Sn wires.

Nb₃Sn has a critical temperature $T_{c0}$ of 18.3 K and upper critical field $B_{c20}$ up to 30 T. An advantage of using high-$C_p$ ribbons on Rutherford cable is that such cable can be used only in those regions of the coils that are at highest magnetic field. This would reduce the impact on quench detection.

Manuscript receipt and acceptance dates will be inserted here.
This work was supported by Fermi Research Alliance, LLC, under contract No. DE-AC02-07CH11359 with the U.S. Department of Energy and US Magnet Development Program.
E. Barzi (Corresponding author), I. Novitsky, D. Turrioni and A.V. Zlobin are with the Fermi National Accelerator Laboratory (Fermilab), Batavia, IL 60510 USA (e-mail: barzi@fnal.gov).
X. Peng and M. Tomsic are with Hyper Tech Research Inc., Columbus, OH 43228.

## II. Experimental Methods

### A. Strand, Cable and High-$C_p$ Ribbon Parameters

Table I shows parameters of the 0.85 mm Restacked-Rod-Processed (RRP) wires produced by Bruker-OST (BOST) for the High-Luminosity Large Hadron Collider Accelerator Upgrade Project (AUP) that were used for this study. The 108/127 stack design notation represents the number of SC bundles in the billet matrix over the total number of restacks.

The high-$C_p$ tapes produced by Hyper Tech for FNAL were of 10 mm width and two different thicknesses, 89 μm and 64 μm, for the $Cu/Gd_2O_3$, and 14 mm wide and 55 μm thick for the $Cu/Gd_2O_2S$. Both ribbons had a 30% fill factor, and Fig. 1 shows a cross section of the $Cu/Gd_2O_3$ tape. For the thinner $Cu/Gd_2O_3$ tape, the enhancement at 4.2 K and 12 T of the heat capacity with respect to a pure Cu tape is by a factor of about 50.

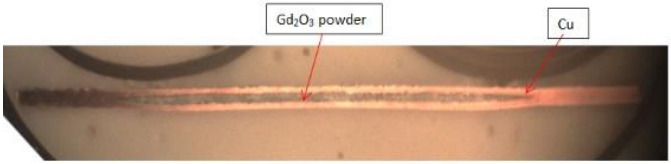

Fig. 1. Cu tape with $Gd_2O_3$ inside, 30% of the cross section is $Gd_2O_3$ (courtesy of Hyper Tech).

TABLE I
$Nb_3Sn$ STRAND PARAMETERS

| Stack Design | Diameter d, mm | $J_c$ (4.2K, 12 T), A/mm$^2$ | $J_c$ (4.2K, 15 T), A/mm$^2$ | Subelement size $D_S$, μm | Twist pitch, mm | Cu fraction λ, % |
|---|---|---|---|---|---|---|
| RRP 108/127 | 0.85 | 2,694 | 1,440 | 55 | 19 | 53.5 |

### B. Strand $I_c$ and MQE Tests

For heat treatment in inert Argon, ~2 m long samples of $Nb_3Sn$ composite wire are wound on grooved cylindrical Ti-alloy (Ti-6Al-4V) barrels and held in place by two removable Ti-alloy end rings. Thermal cycles are performed in three-zone controlled tube furnaces with a 30 cm long temperature homogeneity volume. Calibrated and ungrounded K-type thermocouples are used to monitor the accuracy and homogeneity of the reaction temperatures. After heat treatment, the Ti-alloy end rings are removed from the Ti-alloy barrels and replaced by Cu rings.

Voltage–current ($V$–$I$) characteristics are measured in boiling He at 4.2 K, in a transverse magnetic field. In the wire critical current, or $I_c$, measurement, one pair of voltage taps is used. The pair is placed along the center of the spiral sample 50 cm apart. The $I_c$ is determined from the $V$–$I$ curve using an electrical field criterion of 0.1 μV/cm. Typical $I_c$ measurement uncertainties are within ±1% at 4.2 K and 12 T. The measured $I_c$ value is used when performing the MQE measurement at various normalized transport current ratios $I(B)/I_c(B)$.

To measure the wire MQE, strain gauges are used as heaters. Two heaters are used on the bare sample, and two on the area with the high-$C_p$ tape. Strain gauges WK-09-125BT-350 from Micro-Measurements are glued to the samples using Stycast 2850FT, with the gauge patterns (~4 mm in length and ~1.5 mm in width) centered on the wire sample and their long sides parallel to the wire axis (see Fig. 2, left). After curing of the Stycast, the instrumentation wires are soldered before sample and strain gauge get brushed with a thick layer (~1 mm) of Stycast. A 200 W BPO KEPCO 200-1 power supply provides the excitation voltage to the strain gauge. Using a LabView DAQ program, a 200 μs-long pulse output is generated from the power supply and the voltage across the strain gauge is measured. With the $I_c$ of the sample first measured, a constant bias current below $I_c$ is applied to the sample and heat pulses are fired using the strain gauge. The voltage across the strain gauge and its current are measured with a NI card connected to a NI signal conditioner [2]. A separate voltage detection system monitors the voltage across the sample and shuts down the main power supply if the quench threshold is reached. The typical wait time between heat pulses was 30 s or longer. The characteristic time of Stycast 2850FT for thermal diffusion is much shorter at 8 ms. By gradually increasing the pulse amplitude, i.e. its energy, the minimum energy that induces a quench is defined as the MQE of the sample. The measurement error is given by the difference between the MQE achieved value and the energy provided to the heater at the previous step, at which the sample was not quenching yet. For the wires in this study it was between -0.7% and -3.7%.

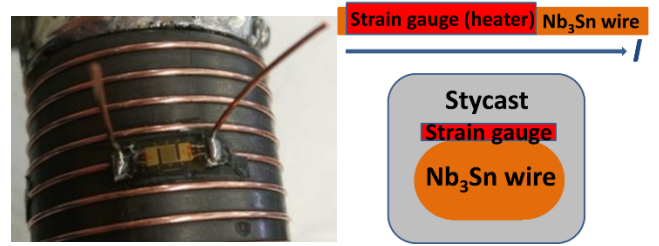

Fig. 2. Strain gauge used as heater mounted on wire (left) and experiment schematic, showing the strain gauge mounted on a flat-rolled wire (right).

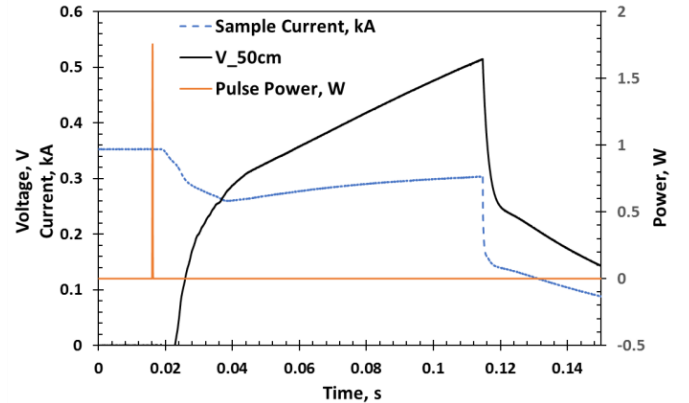

Fig. 3. Voltage signals obtained in the MQE test of the bare rolled $Nb_3Sn$ wire at 15 T and 90% of $I_c$ (15T) with input pulse energy at the MQE value.

To commission the MQE measurements system and achieve reproducibility, several $Nb_3Sn$ wire samples were used [11].

For reproducibility, the most important factor is good thermal contact between the heater and the sample. Since there is no accurate way to measure how much heat from the heater glued to the sample is dissipated into the helium, the following procedure was used. For each sample that was tested, two heaters were mounted on the bare section of the sample and two heaters were mounted on the section outfitted with a high-$C_p$ tape. The complete test is performed for each section by using the heater which provides the smaller *MQE* values on each section. In addition, at least two identical samples are tested for each geometrical configuration, and only the smallest obtained values of *MQE* are used. For instance, in the case of the $Nb_3Sn$ rolled wire soldered to the $Cu/Gd_2O_2S$ tape, three superconducting samples were prepared and tested.

Fig. 3 shows an example of voltage signal during the *MQE* test of a bare sample of $Nb_3Sn$ wire at 15 T and 90% of $I_c$ (15T) with input pulse energy at the *MQE* value. Fig. 4 shows the voltage over sample current ratio obtained in the *MQE* test of the bare $Nb_3Sn$ wire at 15 T and 90% of $I_c$ (15T). During quench development in Fig. 3, it can be seen that the current of the sample power supply, which is operated in constant current mode, reduces to accommodate the larger voltage over the test probe. In Fig. 4, the initial increase of the resistance corresponds to quench development in the wire sample, followed by a lower slope increase due to sample heating. The sharp peak in the *V/I* signal corresponds to the shut-down of the power supply. The resistance subsequently starts slowly decreasing due to sample cooling.

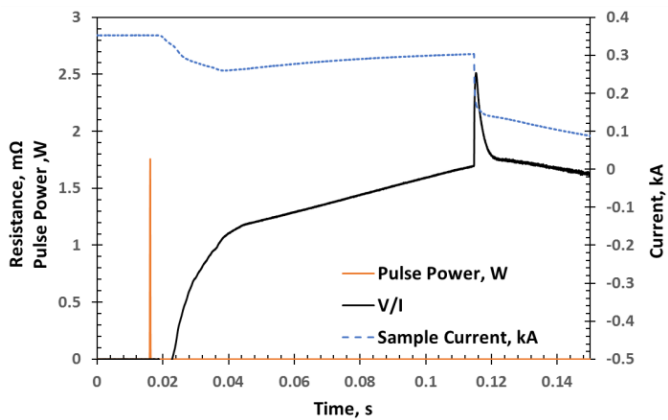

Fig. 4. Voltage over sample current ratios obtained for all voltage tap channels in the *MQE* test of the bare rolled $Nb_3Sn$ wire at 15 T and 90% of $I_c$ (15T).

The $Nb_3Sn$ wire described in Table I was used for these *MQE* wire studies. For some of the geometrical configurations in this study, the wire was flat rolled before reaction from 0.85 mm to 0.71 mm (i.e., 20% deformation) to allow for a better thermal contact with the strain gauge used as heater.

The $Nb_3Sn$ wire was outfitted with the high-$C_p$ tape in three different configurations, after cutting the tape to a width of 2.45 to 2.6 mm:

- Option A – Tape butt-lap wrapped around round wire before heat treatment.
- Options B – Tape placed underneath flat-rolled wire, and either soldered or not to the SC sample after heat treatment.

Option A is shown in Fig. 5 (left), and option B before heat treatment is shown (right) with the high-$C_p$ ribbon underneath.

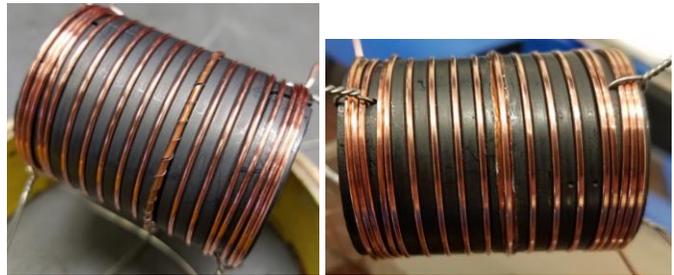

Fig. 5. Options for attaching the high-$C_p$ tape to round wire [11] (Option A, left) and to rolled strands (Option B, right). In both cases the tape was first cut to a width of 2.45 to 2.6 mm.

III. RESULTS AND DISCUSSION

*MQE* results at 15 T for bare round and rolled wires and for options A and B, the latter with the $Cu/Gd_2O_3$ tape both soldered and not soldered to the sample, are shown in Fig. 6 as function of normalized transport current of the $Nb_3Sn$ wire. The results for the round bare wire wrapped with the thinnest (64 μm) of the Hyper Tech $Cu/Gd_2O_3$ tapes (Option A), as shown in Fig. 5 (left) [11], are comparable with the *MQE* achieved for the Hypertech high-$C_p$ wire [10], [12]. Options B were tested using both the thickest and the thinnest $Cu/Gd_2O_3$ tapes, to gauge for differences in stability. The largest *MQE* values were obtained for the Option B configuration with the thickest $Cu/Gd_2O_3$ ribbon soldered to the SC wire. When the tape is not soldered to the SC sample, the *MQE* values of the rolled $Nb_3Sn$ wire are somewhat lower than their bare rolled counterpart. For the former, the 30% reduction of the averaged *MQE* at 60% of the $I_c$ can be explained by the 40% reduction of transverse thermal conductivity [13] of the wire-ribbon assemblage, using thermal conductivities for the Cu and the $Gd_2O_3$ respectively of 160 W/(mK) and 6.2 W/(mK) [11]. Another observation is that the *MQE* obtained for bare flat-rolled wires is higher than that of round ones. This can be explained with the increased heat transfer surface, resulting in higher heat transfer towards the Stycast and therefore in a lower increase in the wire temperature.

*MQE* results at 14 T for bare round and rolled wires and for options A and B, the latter with the $Cu/Gd_2O_2S$ tape soldered to the sample, are shown in Fig. 7 as function of normalized transport current of the $Nb_3Sn$ wire.

Fig. 8 shows the gain, or the *MQE* ratio between the *MQE* obtained for the $Nb_3Sn$ wire with the high-$C_p$ ribbon and that without, as function of the wire normalized transport current. The wire samples soldered with $Cu/Gd_2O_3$ tape have *MQE* values between 1.8 and 3.3 times higher than the bare specimen at up to 80% of $I_c$. The wire samples outfitted with $Cu/Gd_2O_2S$ tape have *MQE* values between 1.7 and 2.1 times higher than the bare specimen at up to 95% of $I_c$.

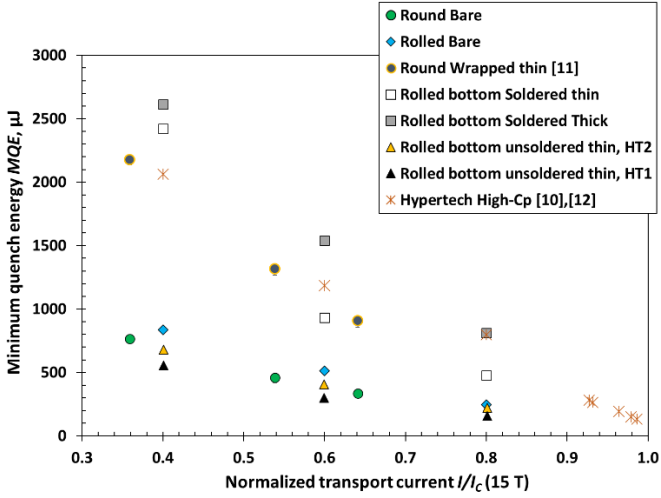

Fig. 6. *MQE* results at 15 T for bare round and rolled wires and for options A and B, the latter with the Cu/Gd$_2$O$_3$ tape both soldered and not soldered to the sample, as function of normalized transport current for the AUP wire. "HT1" and "HT2" in the legend indicate tests performed on the same sample with two different heaters.

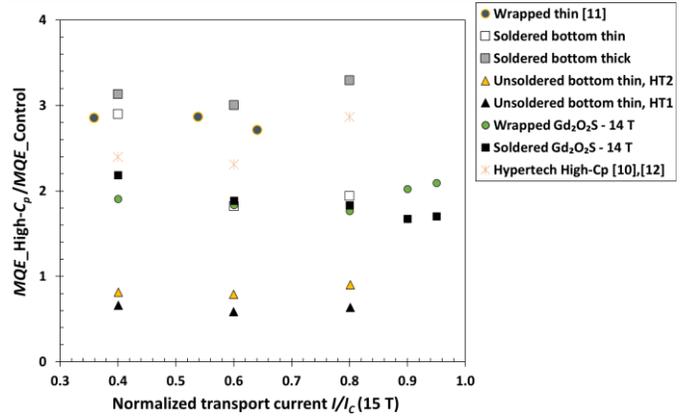

Fig. 8. Gain of outfitting a wire with high-$C_p$ tape, or the *MQE* ratio between the *MQE* obtained for the Nb$_3$Sn with the ribbon and that without, as function of normalized transport current. For the Hyper Tech wire, the gain was plotted as *MQE* of the high-$C_p$ wire over the *MQE* of the control standard wire. "HT1" and "HT2" in the legend indicate tests performed on the same sample with two different heaters.

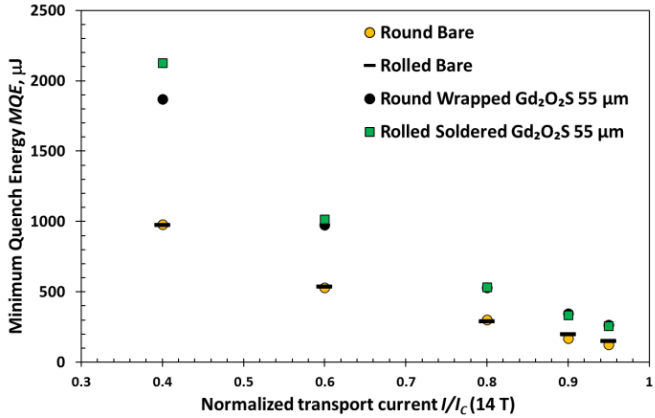

Fig. 7. *MQE* results at 14 T for bare round and rolled Nb$_3$Sn wires and for options A and B, the latter with the Cu/Gd$_2$O$_2$S tape soldered to the sample, as function of normalized transport current for the AUP wire.

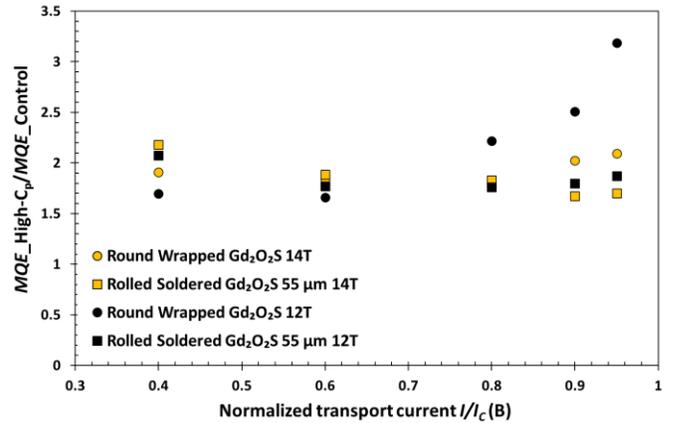

Fig. 9. Gain values of Nb$_3$Sn wire wrapped or soldered with the Cu/Gd$_2$O$_2$S tape as function of normalized transport current at various magnetic fields.

When looking for correlations between wire and cable tests, a possible representation of the Nb$_3$Sn cable test with high-$C_p$ tape is that of the ribbon soldered to the SC wire, in the assumption that ribbon sintering to the bare cable faces will occur during coil reaction. Therefore, future Nb$_3$Sn cable test results with high-$C_p$ tape on both sides will be compared with the wire outfitted with the soldered tape. If a consistent correlation will be achieved, this will also confirm sintering of the tape on the cable faces during high temperature reaction.

To check for magnetic field dependence, the Nb$_3$Sn wire gain values when outfitted with Cu/Gd$_2$O$_2$S tape were measured at several magnetic fields. Fig. 9 shows the results obtained so far. Within the limited statistics, there does not appear to be any magnetic field dependence on gain.

## IV. Conclusions

A major focus of Nb$_3$Sn accelerator magnets is on significantly reducing or eliminating training. Increasing the conductor specific heat will lead to shorter training with substantial savings in machines commissioning costs. Samples of Hyper Tech high-$C_p$ Cu/Gd$_2$O$_3$ and Cu/Gd$_2$O$_2$S tapes were used to measure and compare the *MQE* of Nb$_3$Sn wires outfitted with this tape. A gain in *MQE* of 1.7 to 3.3 was found at magnetic fields between 12 T and 14 T. For an even more precise measure of the effect of a high-$C_p$ tape, the *MQE* results herein obtained will be compared in the close future with the *MQE* of a pure Cu tape wrapped around or soldered to a superconducting wire.

Next, Nb$_3$Sn cable test results with Cu/Gd$_2$O$_2$S tape on both sides will be compared with the wire outfitted with the soldered tape. If a consistent correlation will be achieved, this will also confirm sintering of the tape on the cable faces during high temperature reaction.

## References

[1] Stoynev, S.E.; Riemer, K.; Zlobin, A.; Ambrosio, G.; Ferracin, P.; Sabbi, G.; Wanderer, P., "Analysis of Nb$_3$Sn Accelerator Magnet Training," IEEE Trans. Appl. Supercond. 2019, 29, 1–6, doi:10.1109/tasc.2019.2895554.

[2] Barzi, E,; Novitski, I.; Rusy, A.; Turrioni, D.; Zlobin, A.V.; Peng. X.; Tomsic, M., "Test of Superconducting Wires and Rutherford Cables with High Specific Heat", IEEE Trans. Appl. Supercond., V. 31, no. 5, pp. 1-8, 2021, Art no. 6000508, doi: 10.1109/TASC.2021.3069047.

[3] Meuris, C., "Thermal Stability of Superconductors", J de Physique (1984) 45 503-510.

[4] Martinelli, A.P. and Wipf, S.L., "Investigation of Cryogenic Stability," Proc 1972 Appl Supercond Conf IEEE, New York, USA (1972) 325-330.

[5] Wilson, M.N. and Iwasa, Y., "Stability of Superconductors Against Localized Disturbances of Limited Magnitude," Cryogenics (1978) 18 17-25.

[6] Buznikov, N.A. and Pukhov, A.A., "Analytical Method to Calculate the Quench Energy of a Superconductor Carrying a Transport Current," Cryogenics 1996, 37, 7 547-553.

[7] Keilin, V.E. and Romanovsky, V.R. "The Dimensionless Analysis of the Stability of Composite Superconductors with Respect to Thermal Disturbances", Cryogenics (1982) 22 313-317.

[8] Gurevich, A.Vl., Mints, R.G. and Pukhov, A.A., "Quench Energies of Composite Superconductors", Cryogenics (1989) 29 199-190.

[9] Hancox, R., "Enthalpy Stabilized Superconducting Magnets," IEEE Trans. Magn. 1968, 4, 486–488, doi:10.1109/tmag.1968.1066271.

[10] Xu, X.; Li, P.; Zlobin, A.V.; Peng, X., "Improvement of Stability of Nb$_3$Sn Superconductors by Introducing High Specific Heat Substances," Supercond. Sci. Technol. 2018, 31, 03LT02, doi:10.1088/1361-6668/aaa5de.

[11] Barzi, E.; Berritta, F.; Turrioni, D.; Zlobin, A.V., "Heat Diffusion in High-$C_p$ Nb$_3$Sn Composite Superconducting Wires," Instruments 2020, 4, 28.

[12] Xu, X.; Zlobin, A.; Peng, X.; Li, P., "Development and Study of Nb$_3$Sn Wires with High Specific Heat," IEEE Trans. Appl. Supercond. 2019, 29, 1–4, doi:10.1109/tasc.2019.2892325.

[13] Breschi, M.; Trevisani, L.; Bottura, L.; Devred, A.; and Trillaud, F., "Effects of the Nb$_3$Sn Wire Cross Section Configuration on the Thermal Stability Performance," IEEE Trans. Appl. Supercond., V. 19, No. 3, June 2009, pp. 2432-2436.